# U-SWIFT: A Unified Surface Wave Inversion Framework with Transformer via Normalization of Dispersion Curves

Tianjian Cheng[1,*], Hongrui Xu[1], Jiayu Feng[1], Xiongyu Hu[2], and Chaofan Yao[3]

[1]Faculty of Geoscience and Engineering, Southwest Jiaotong University, Chengdu, China

[2]Key Laboratory of Transportation Tunnel Engineering, Ministry of Education, Southwest Jiaotong University, Chengdu, China

[3]State Key Laboratory of Intelligent Geotechnics and Tunnelling, Southwest Jiaotong University, Chengdu, China

* E-mail: tjcheng.ok@163.com

Corresponding author: Hongrui Xu (hongrui_xu@swjtu.edu.cn)

**Abstract**

Deep learning is an increasingly popular approach for inverting surface wave dispersion curves to obtain $V_S$ profiles. However, its generalizability is constrained by the depth and velocity scales of training data. We propose a unified deep learning framework that overcomes this limitation via normalization of dispersion curves. By leveraging the scaling properties of dispersion curves, our approach enables a single, pre-trained model to predict $V_S$ profiles across diverse scales, from shallow subsurface (e.g., < 10 m depth) to crustal levels. The framework incorporates a novel transformer-based model to handle variable-length dispersion curves and removes tedious manual parameterization. Results from synthetic and field data demonstrate that it delivers rapid and robust inversions with uncertainty estimates. This work provides an efficient inversion approach applicable to a wide spectrum of applications, from near-surface engineering to crustal imaging. The framework establishes a paradigm for developing scale-invariant deep learning models in geophysical inversion.

# 1 Introduction

The inversion of surface wave dispersion curves to obtain subsurface shear wave velocity ($V_S$) profiles is a fundamental approach in geophysical subsurface imaging across various depth scales (Aki & Richards, 2009). It has been widely implemented in near-surface engineering applications (Foti et al., 2011; Gohil et al., 2024; Gosselin et al., 2022; Lin et al., 2004) and deep crustal scientific research (Bensen et al., 2009; Fang et al., 2015; Shen & Ritzwoller, 2016; Xu et al., 2024). This inverse problem is commonly addressed using two primary classes of methods: gradient-based local optimization methods and stochastic global optimization methods. Gradient-based methods (Wu et al., 2020; Yanovskaya & Kozhevnikov, 2003; Yin et al., 2020) use the gradients to accelerate the optimization process but are sensitive to initial models and prone to converging on local minima. In contrast, stochastic methods, such as simulated annealing algorithm (Beaty et al., 2002; Pei et al., 2007), Markov chain Monte Carlo algorithm (Berg et al., 2018; Bodin et al., 2012; Saifuddin et al., 2018), genetic algorithm (Dal Moro et al., 2007; Lomax & Snieder, 1995; Yamanaka & Ishida, 1996), and neighborhood algorithm (Wathelet, 2008; Wathelet et al., 2004) can escape local minima by exploring the solution space more broadly through random sampling. However, these methods are computationally intensive.

With the rapid development of artificial intelligence, deep learning using neural networks has become an increasingly popular approach for inverting surface wave dispersion curves. Compared to conventional gradient-based or stochastic inversion methods, deep learning utilizes training data to learn the mapping function from dispersion data to $V_S$ profiles. Once trained, the model can ideally produce results in a single forward pass, eliminating the need for an iterative optimization process. This characteristic gives deep learning a significant advantage over traditional methods in terms of computational efficiency.

Early work explored the potential of fully-connected neural networks (ANNs) and convolutional neural networks (CNNs) to invert surface wave dispersion curves at the crustal scale (Hu et al., 2020; Luo et al., 2022). Subsequent advancements have refined these methods by improving accuracy and efficiency. For example, a weighted mean squared relative error is introduced to improve model performance (Chen et al., 2022). End-to-end CNNs are developed to bypass the need for manual dispersion curve picking (Cho et al., 2024) and shallow neural networks are developed to reduce training time (Yang et al., 2022). Attention has also been given to uncertainty quantification, with the use of mixture density networks to output probability distributions of $V_S$ (Keil & Wassermann, 2023), or a combination of Monte Carlo simulations and ANNs to project errors onto the final model (Yablokov et al., 2023). Furthermore, to better constrain the inversion, data fusion techniques have been employed, such as incorporating P-wave velocity and density into network inputs (Chen et al., 2024) or jointly inverting surface-wave dispersions and receiver functions using multi-network architectures (Wang et al., 2024).

Despite recent advancements, current deep learning models have limitations when predicting $V_S$ profiles across diverse depth and velocity scales. This challenge arises because information regarding specific scales is implicitly encoded in the training data. Consequently, when a pre-trained network encounters an inversion problem with scales different from its training data, it is likely to produce inaccurate results. This shortcoming necessitates a complete repetition of the data generation and training cycle for each new set of target scales. Furthermore, these models require a fixed number of input features, rendering them unable to adapt to dispersion curves of varying lengths. These limitations in adaptability significantly impede the widespread practical application of deep learning models for dispersion curve inversion.

In this study, we propose a unified deep learning framework for the inversion of fundamental-mode Rayleigh wave dispersion curves via normalization of dispersion curves, which enables a single training procedure to accommodate diverse depth and velocity scales. The normalization is achieved by leveraging the scaling properties of dispersion curves. These properties establish that two dispersion curves are translational equivalents when their layer parameters satisfy certain proportionality relations (Aimar et al., 2024; Maraschini et al., 2011; Socco & Strobbia, 2004). We begin by reviewing these scaling properties for one-dimensional layered earth models. Subsequently, each component of the proposed framework is presented and discussed in detail. Within this framework, we develop a specific transformer model designed to handle dispersion curves of varying lengths. Finally, the efficacy of this singly trained model and its ability to provide robust uncertainty estimates are demonstrated using both theoretical and observed dispersion curves that feature distinctly different depth and velocity scales.

## 2 Data and Methods

### 2.1 The Unified Framework

For elastic horizontally layered models, the dispersion equation is governed by three categories of layer parameters: mass density, layer thickness, and body wave velocities. Each parameter class has a corresponding scaling property that dictates how the dispersion equation transforms when all parameters within that class are scaled by a uniform factor across all layers. To visualize the scaling properties, Figures 1a-1c illustrate the velocity and density profiles for a base model (Xia et al., 1999) and three scaled models. Each scaled model is generated by scaling only one parameter class by a factor greater than one. Figure 1d shows the fundamental-mode Rayleigh-wave dispersion curves for all four models on a log-log plot of frequency versus Rayleigh-wave phase velocity. Notably, all curves exhibit identical shapes. Relative to the base model, the dispersion curve for the density-scaled model remains unchanged while the dispersion curves for the thickness-scaled and velocity-scaled models shift horizontally (toward lower frequencies) and diagonally (toward higher frequencies and velocities), respectively.

Based on the thickness and velocity scaling properties, a Normalized Dispersion Curve Space (NDCS) can be constructed. This space is defined as the set of all possible dispersion curves generated from velocity profiles with a fixed half-space depth $H_0$ and body wave velocities ($V_{S0}$, $V_{P0}$). Any dispersion curve, derived from a velocity profile with an arbitrary half-space depth $H_{hs}$ and S-wave velocity $V_{S,hs}$, can be normalized to an equivalent curve in the NDCS.

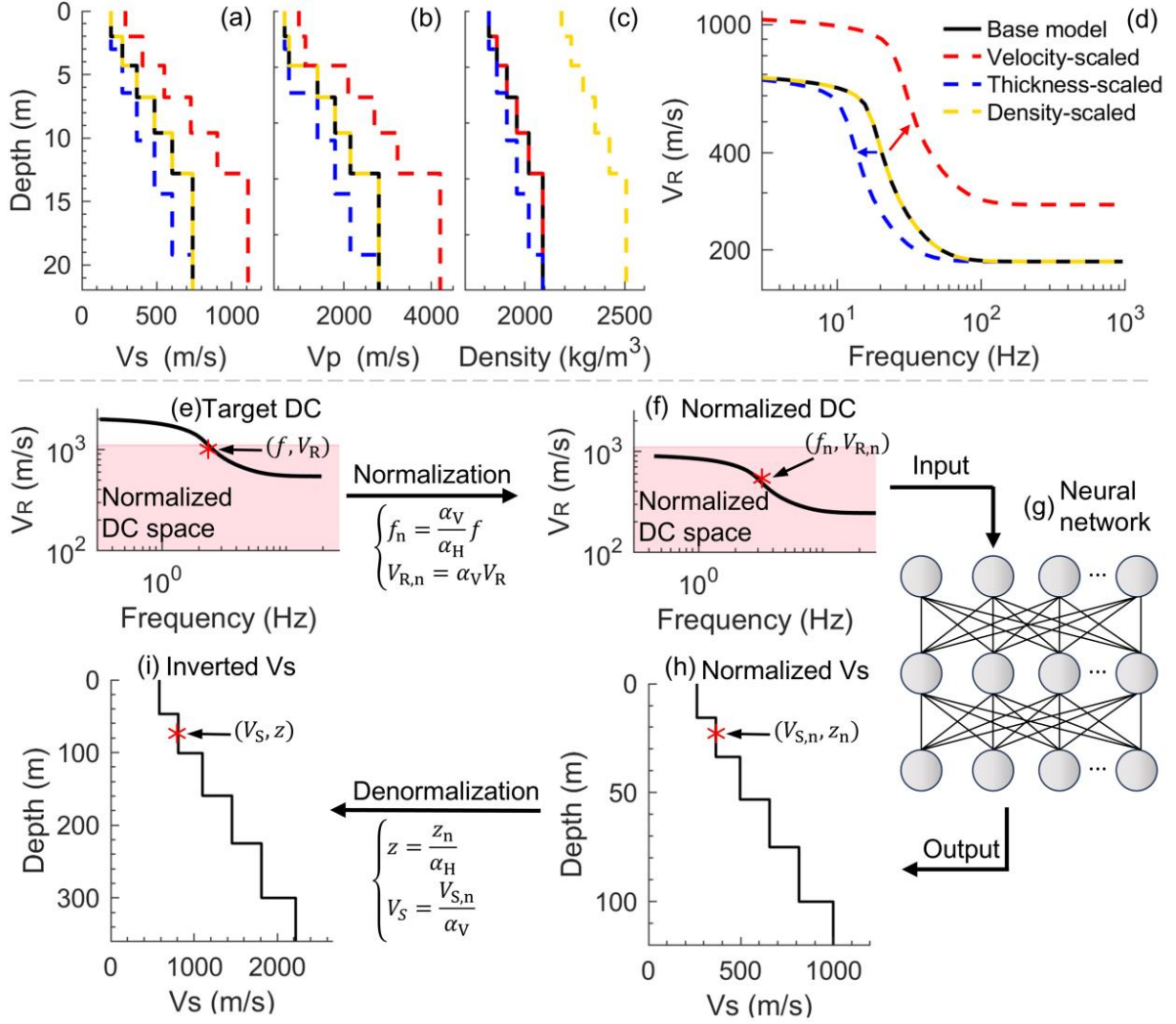

**Figure 1.** (a-c) $V_S$, $V_P$ and density of the base, velocity-scaled, thickness-scaled and density-scaled models; (d) the Rayleigh-wave fundamental dispersion curves of the four models. The dispersion curves (DCs) of the scaled models are either identical to or a translation of the DC of the base model. Parameters of the four models are shown in Table S1-S4 in Supporting Information; (e-i) The unified framework for dispersion inversion with deep learning using the scaling properties.

This normalization is achieved using a depth scaling factor $\alpha_H$ and velocity scaling factor $\alpha_V$, which are calculated as follows:

$$\begin{cases} \alpha_H = \dfrac{H_0}{H_{hs}} \\ \alpha_V = \dfrac{V_{S0}}{V_{S,hs}} \end{cases} \quad (1)$$

Figures 1e-1i show the unified framework for dispersion inversion with deep learning models, which leverages the scaling properties and the concept of NDCS. The process begins by normalizing a target dispersion curve into a predefined NDCS. Concretely, any point $(f, V_R)$ on the target dispersion curve is transformed into a corresponding point ($f_n$, $V_{R,n}$) on the normalized curve according to the following equations:

$$\begin{cases} f_n = \dfrac{\alpha_V}{\alpha_H} f \\ V_{R,n} = \alpha_V V_R \end{cases} \quad (2)$$

A detailed proof for this transformation is provided in Text S1 and Figure S1 in Supporting Information S1. The normalized curve is then processed by neural networks, trained exclusively on dispersion curves within the NDCS, to produce a normalized $V_S$ profile. Finally, this normalized profile is denormalized using the scaling factors to yield the final inverted $V_S$ profile for the target dispersion curve. Concretely, normalized shear wave velocity $V_{S,n}$ at normalized depth $z_n$ is transformed into inverted shear wave velocities $V_S$ at depth $z$ according to the following equations:

$$\begin{cases} z = \dfrac{z_n}{\alpha_H} \\ V_S = \dfrac{V_{S,n}}{\alpha_V} \end{cases} \quad (3)$$

By training in a normalized domain, the deep learning model learns a universal, scale-invariant mapping between the dispersion curves and corresponding subsurface profiles. Consequently, a single, pre-trained model can be universally applied to dispersion inversion problems across diverse depth and velocity scales, eliminating the need for case-specific data generation and retraining cycles. Therefore, this framework significantly enhances computational efficiency. Furthermore, this framework improves learning dynamics by operating within a constrained and consistent input space (Bishop, 2006; Ian Goodfellow et al., 2016; Ioffe & Szegedy, 2015). The model can focus its capacity on resolving the intrinsic complexities of the problem rather than learning superficial, scale-related variations. This focus often results in a more robust and accurate model. In essence, this unified framework decouples the underlying physical relationships from the specific scales of an inversion problem, yielding a flexible, efficient, and broadly applicable deep learning solution.

2.2 Dataset

To show the effectiveness of the unified framework, we developed a pipeline to predict $V_S$ profiles from fundamental-mode Rayleigh-wave dispersion curves. An NDCS with a half-space depth of 100 m and $V_S$ of 1000 m/s was selected for this pipeline. A dataset for training and testing was generated by sampling synthetic velocity profiles within this NDCS and computing their corresponding dispersion curves. Since dispersion inversion is inherently an ill-posed optimization problem with significant non-uniqueness, geophysical constraints are commonly employed to mitigate non-uniqueness in traditional inversion methods. Likewise, we incorporated two geophysical constraints into the model sampling process. The first constraint prohibits low velocity zones (LVZs), requiring velocity to increase monotonically with depth. This is a common constraint in dispersion inversion, as LVZs are relatively rare and their inclusion significantly increases the non-uniqueness of the solution (Haney & Tsai, 2017; Luo et al., 2007; Wathelet, 2005). Second, a constant density of 2000 kg/m$^3$ and Poisson's ratio of 0.33 were set across all velocity profiles. This simplification is justified because these parameters have a substantially smaller influence on Rayleigh wave dispersion than $V_S$ and layer thickness (Gosselin et al., 2022). Consequently, their influence on the inverted $V_S$ profile is limited. Since the primary objective is to demonstrate the framework's effectiveness, these simplifications provide a reasonable and robust baseline for evaluation, with refinements to be explored in future work.

Next, the 100-m-thick stratum above the half-space was discretized into 100 one-m-thick layers. Consequently, a velocity profile within the NDCS is defined solely by the $V_S$ values of these 100 layers. This parameterization scheme transforms the inversion problem from one with a variable number of unknown parameters due to varying layer numbers into one with a constant number of unknown parameters (100 $V_S$ values), while still providing high vertical resolution. This standardization is crucial, as neural networks require fixed-size output vectors. As a result, the network can be designed to output a high-resolution $V_S$ profile without needing the number of layers as an a priori input, which is often unknown in practice.

$V_S$ values inverted from field dispersion curves rarely fall below 1/20th (50 m/s) of the half-space velocity, so we set the lower bound for our synthetic $V_S$ profiles to this same ratio. To mitigate sampling bias and efficiently produce $V_S$ profiles satisfying the no-LVZ constraint, we propose the Randomized Layer Assignment Method to sample $V_S$ values for the 100 discretized layers. This method is initiated by uniformly sampling the $V_S$ values of the top and bottom layers between 50 and 1000 m/s. Subsequently, the remaining 98 layers are assigned $V_S$ values iteratively. In each step, an unassigned layer is selected at random, and its $V_S$ is sampled uniformly from the interval defined by the $V_S$ values of the nearest assigned layers above and below it. This method thoroughly explores the parameter space and ensures that each layer has an approximately equal probability of attaining any velocity between 50 and 1000 m/s, which is crucial for enhancing deep learning models' generalizability. More information about the sampling method is provided in Text S2 and Figure S2 in Supporting Information S1. Using this method, we generated one million $V_S$ profiles for the training dataset and one hundred thousand $V_S$ profiles for the testing dataset.

2.3 Model Structure

We propose a deep learning model called the Point-wise Additive Dispersion Inversion Transformer (PADIT) to invert variable-length dispersion curves within the defined NDCS, as shown in Figure 2a. The model is composed of three main blocks: the embedding layer, the encoder block with six identical encoder layers, and the regression head, with their structures detailed in Figures 2b-2d. The embedding layer is a multilayer perceptron (MLP) that maps each point on the normalized dispersion curve into a high-dimensional feature vector. Positional encoding is then added to each vector to inject information about its sequential position. Following this, the complete set of feature vectors is processed sequentially by six encoder layers. The output vectors are aggregated via sum pooling to yield a single, fixed-size global descriptor regardless of the input curve's length. Finally, this global descriptor is processed by the regression head to produce the $V_S$ values for the 100 one-meter-thick layers.

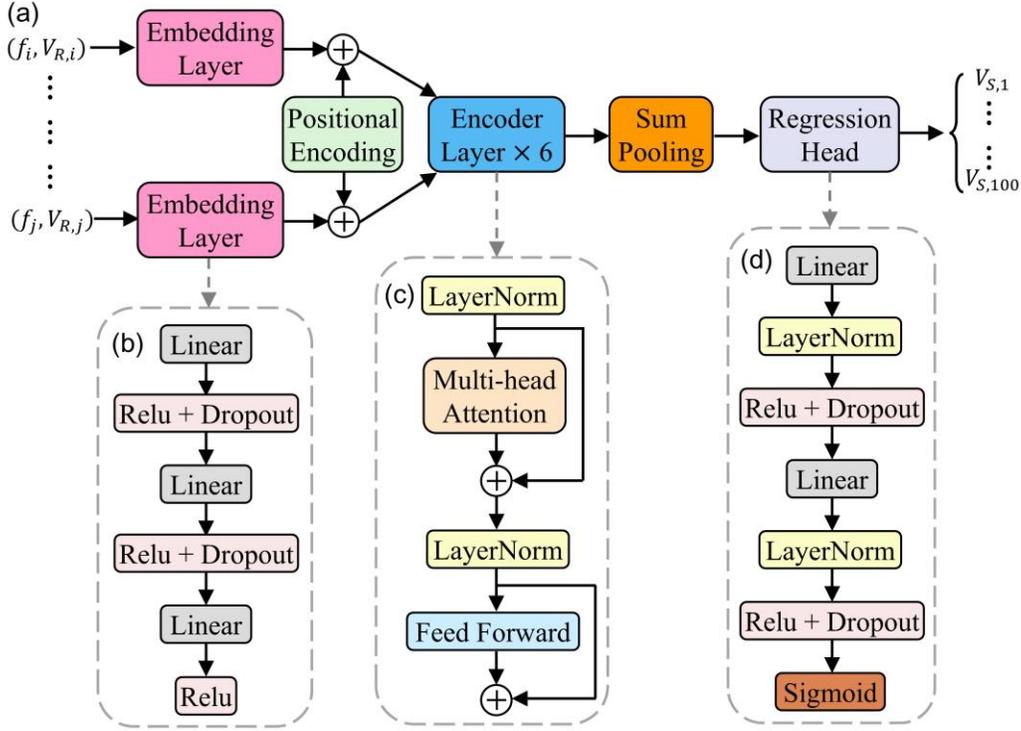

**Figure 2.** (a) Overall Architecture of the proposed model; (b-d) the detailed structures of the embedding layer, encoder layer, and regression head, respectively. The plus sign ($\oplus$) denotes element-wise addition.

A key data augmentation technique we employed is to crop the dispersion curves into sections of varying lengths and frequency ranges before training. This strategy forces the model to learn the underlying physical relationships from partial data, which greatly improves its generalizability. To balance errors across shallow low-velocity and deep high-velocity sections of the inverted $V_S$ profiles, the mean absolute percentage error (MAPE) was selected as the loss function. The AdamW optimizer, combined with the cosine annealing technique, is used to optimize the training process. Finally, the model reaches a low MAPE of 1% on the training set and 2% on the testing set (Figure S3 in Supporting Information S1), which can be visually confirmed by the strong agreement between target and predicted dispersion curves and $V_S$ profiles (Figure S4-S6 in Supporting Information S1).

2.4 The Scaling Factors

A critical step for the successful application of the unified framework is the estimation of reasonable scaling factors, which requires knowledge of the real-world half-space depth $H_{hs}$ and shear wave velocity $V_{S,hs}$. The two parameters are usually unknown a priori, but their ranges can be constrained by the maximum resolved wavelength $\lambda_{max}$ and corresponding phase velocity $V_{R,max}$ of the field dispersion curve (Cox & Teague, 2016; Garofalo et al., 2016; Richart et al., 1970). For initial inversion, we have found empirically that reasonable ranges for $H_{hs}$ and $V_{S,hs}$ can be set as broad as [0.2 $\lambda_{max}$, $\lambda_{max}$] and [1.07 $V_{R,max}$, 3.0 $V_{R,max}$], respectively. Notably, if the field curve exhibits a horizontal asymptote in the low-frequency range, the ranges of $H_{hs}$ and the upper bound of $V_{S,hs}$ could be significantly smaller.

Once the ranges are determined, multiple pairs of ($H_{hs}$, $V_{S,hs}$) are sampled systematically and fed into the framework to generate a corresponding set of inverted $V_S$ profiles. The inverted $V_S$ profiles are ranked based on the misfit (Wathelet et al., 2004) between their dispersion curves and the field curve. Unlike traditional iterative inversion methods, this process can be greatly accelerated by parallel computing because the inversions for each ($H_{hs}$, $V_{S,hs}$) pair are independent. Ideally, the dispersion curves of top-ranked models should closely match the field curve. Values of $H_{hs}$ and $V_{S,hs}$ from these models can be used to derive refined ranges if a subsequent, more accurate inversion is desired. However, if this process yields few valid solutions, the initial search ranges for the half-space parameters are likely inadequate and should be broadened.

## 3 Results

3.1 Assessment with Variable-Length Synthetic Data

To evaluate the model's ability to handle dispersion curves of varying lengths, we performed a synthetic test using a theoretical $V_S$ profile. The complete theoretical dispersion curve corresponding to this profile was truncated to create three separate inputs of different lengths (long, medium, and short). Each was fed into the deep learning framework for inversion. We assigned each data point a standard deviation equal to 1% of its phase velocity and only considered results with a misfit below 1 as valid. As shown in Figure 3, the best-fit dispersion curves and their corresponding inverted $V_S$ profiles from all three inputs show close agreement with the theoretical ground truth. This demonstrates that the framework can robustly and accurately invert dispersion curves of varying lengths.

An important observation is the increase in both the number and diversity of valid solutions as the target dispersion curve becomes shorter, which is visually confirmed by the broadening distribution of the valid $V_S$ profiles and their corresponding dispersion curves in Figure 3. This finding is consistent with the widely-acknowledged principle that a reduction in the frequency range of the target dispersion curve increases the non-uniqueness of the inversion. The increase in non-uniqueness is mild when transitioning from the long to the medium-length curve, as the medium-length curve still exhibits asymptotic behavior at both the low- and high-frequency ends. However, the non-uniqueness increases dramatically for the short curve because the asymptotic behaviors at both ends vanish. This results in a much wider distribution of $V_S$ values in the shallowest and deepest layers. This phenomenon suggests that the deep learning model has the capability to capture the physical non-uniqueness inherent in the inversion problem. It implies that our framework could be used as a tool not only for inverting a single best-fit $V_S$ profile but also for characterizing the uncertainty of the solution.

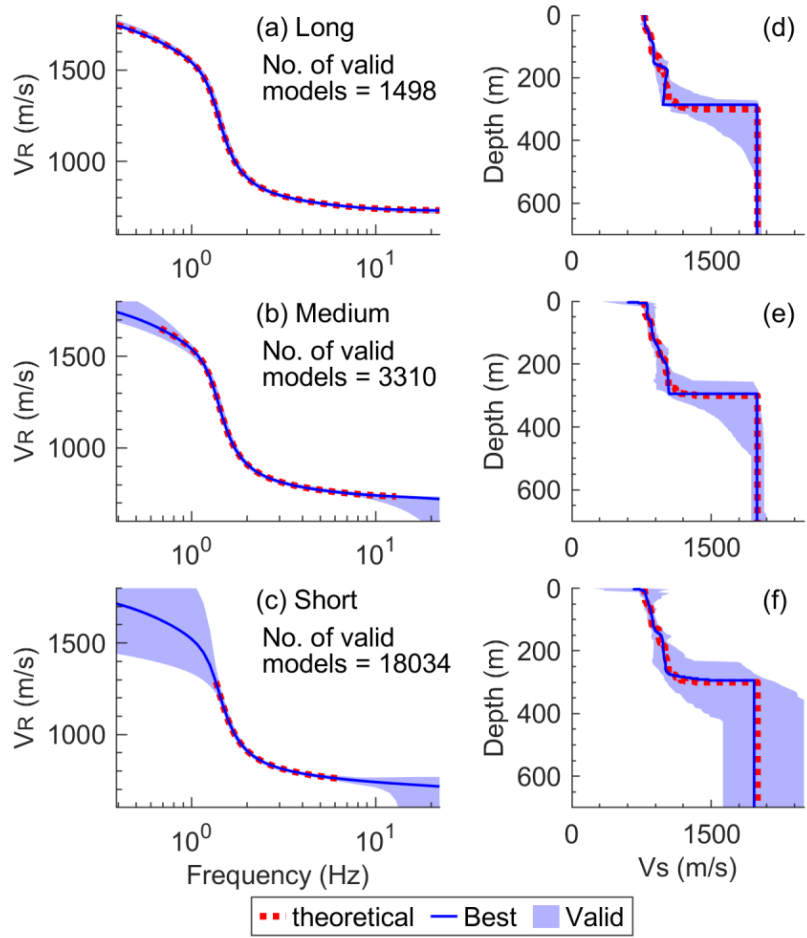

**Figure 3.** Inversion results for dispersion curves of varying lengths. (a-c) Comparison of the best-fit (blue line) and valid (light blue area) dispersion curves against the actual curve (red dotted line) for inputs of long, medium, and short length. (d-f) The corresponding inverted $V_S$ profiles. The valid models are all solutions with a misfit below 1.

3.2 Application to Multi-Scale Field Datasets

We compared the inversion results from our framework with those from a traditional Markov chain Monte Carlo (MCMC) approach using dispersion curves from three sites with distinct depth and velocity scales: the western Bohai Bay Basin (Xu et al., 2024), the Balikun Basin (Luo et al., 2018), and the Boise Hydrogeophysical Research Site (BHRS; Mi et al., 2020; Xu et al., 2022). As shown in Figure 4a-c, the measured dispersion curves cover substantially different frequency ranges of 0.025-0.2 Hz, 2-24 Hz, and 20-55 Hz, respectively. For all three sites, the dispersion curves forward-modeled from both the AI-predicted and the MCMC-derived $V_S$ profiles show excellent agreement with the observed data. The inverted $V_S$ profiles from both methods exhibit strong consistency across most depth ranges (Figure 4d-4f).

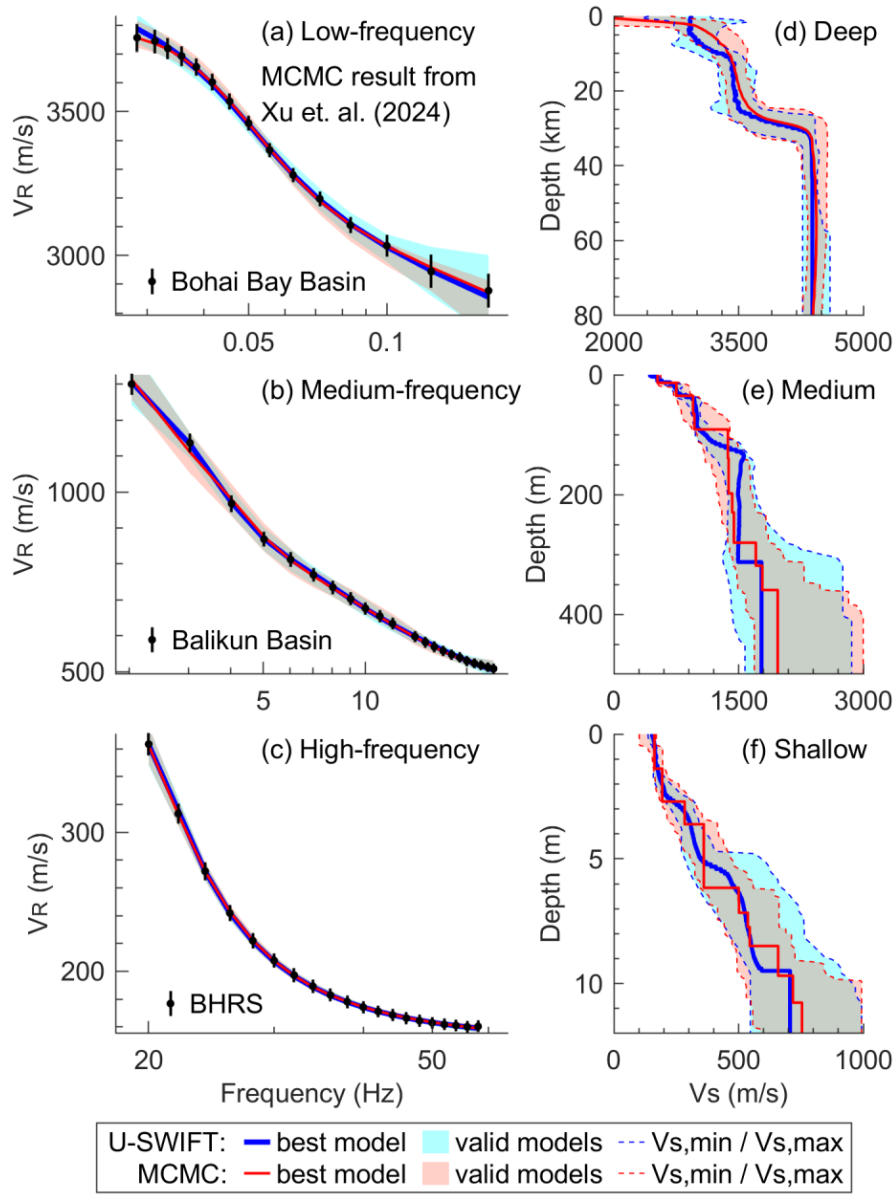

**Figure 4.** Comparison of inversion results from our framework (U-SWIFT) and the traditional Markov chain Monte Carlo (MCMC) method for three field datasets at different scales. (a-c) The best-fit dispersion curves (solid lines) and spaces of valid models (shaded areas) for both U-SWIFT (blue) and MCMC (red) compared against the observed data (black error bars). (d-f) The corresponding inverted $V_S$ profiles. The dashed blue lines indicate the minimum and maximum velocity bounds for the valid solutions. For both methods, valid models are defined as solutions with a misfit below 1. The MCMC result for the Bohai Bay Basin is from Xu et al. (2024).

For the deep crustal case, both methods resolve a pronounced velocity increase near the Moho interface and show a corresponding increase in uncertainty at these depths (Figure 4d). While some discrepancies exist at shallow depths in the deep crustal model, mainly due to the low sensitivity of surface waves below 0.2 Hz to the uppermost few kilometers, the overall agreement between the two methods remains excellent. Similarly, in the two shallower cases, the uncertainty estimates from both methods gradually increase with depth (Figures 4e-f). These findings confirm that our framework delivers robust inversion results and uncertainty assessments that are consistent with traditional methods across a wide range of velocity and depth scales.

## 4 Discussion and Conclusion

In this study, we proposed a unified deep learning framework for the inversion of surface wave dispersion curves, designed to overcome critical limitations of previous deep learning models. Results from both synthetic and field data demonstrate the framework's accuracy, flexibility, and robustness. The success of the framework stems from four key characteristics that significantly enhance its practical applicability.

First, the framework can handle inversion problems across diverse depth and velocity scales with a single, pre-trained deep learning model. Existing deep learning models are typically constrained by scales implicit in their training data, requiring specialized and computationally intensive retraining to invert dispersion curves with different investigation depths or velocity ranges. By leveraging the scaling properties of dispersion curves, our method normalizes any field dispersion curve into a Normalized Dispersion Curve Space (NDCS). This decouples the intrinsic physical relationship between dispersion and $V_S$ profiles from the specific scale of the problem. As demonstrated by the inversion of three field datasets from shallow near-surface to deep crustal levels, our framework exhibits strong generalizability, making it a highly efficient tool for inverting any field dispersion curve.

Second, the proposed Point-wise Additive Dispersion Inversion Transformer (PADIT) model is specifically designed to accommodate dispersion curves of varying lengths. In practice, the length of field dispersion curves can differ significantly due to different site conditions, acquisition parameters, and processing methods. Our model does not require a fixed-size input because it processes each data point on the dispersion curve individually before aggregating the feature vectors. The synthetic test shows that the model can accurately invert long, medium-length, and short segments of a theoretical dispersion curve, confirming its robustness to input data length.

Third, the synthetic test revealed that the framework's output reflects the inherent non-uniqueness of the inversion problem. As the input curve becomes shorter, the range of valid inverted profiles broadens at the depths corresponding to the missing frequency ranges. The comparison with the MCMC approach at three real-world sites further demonstrates that the AI model not only produces reliable multi-scale $V_S$ models but also yields meaningful uncertainty estimates across all depths.

Fourth, our framework simplifies the inversion workflow by eliminating the need for tedious manual parameterization. Traditional inversion methods require users to make subjective choices, such as defining the number of subsurface layers and setting specific velocity constraints (Vantassel & Cox, 2020). The MCMC approach used for the deep crustal case further illustrates this complexity, requiring the lithosphere to be subdivided into sedimentary, crystalline crust, and uppermost mantle layers, along with strict prior constraints based on reference models (Xu et al., 2024). These choices not only substantially influence the outcomes but also demand extensive region-specific prior information. Our approach only requires broad estimates for the half-space depth and $V_S$, which can be estimated from the dispersion curve itself. This framework, combined with parallel computing, significantly accelerates the inversion process, making it feasible to obtain near-real-time results in the field.

While the current model was developed under specific assumptions to establish a robust baseline, our findings underscore the potential of this unified deep learning framework to significantly advance surface wave dispersion inversion and potentially other geophysical inverse problems. Its ability to deliver rapid and reliable results offers valuable insights for a wide spectrum of applications, from near-surface engineering to deep crustal imaging. This work therefore contributes a powerful and efficient paradigm, providing a solid foundation for the next generation of deep learning solutions in geophysical inversion.

**Conflict of Interest**

The authors declare no conflicts of interest relevant to this study.

**Data Availability Statement**

Data and code are available at https://github.com/CTJ-UT/U-SWIFT-A-Unified-Surface-Wave-Inversion-Framework-with-Transformer.

**Acknowledgments**

The research was supported by the National Key R&D Program of China (Grant No. 2024YFB2606101), the China Postdoctoral Science Foundation (Grant No. 2023M732912), and the Fundamental Research Funds for the Central Universities (Grant No. A0920502052401-210).

Supporting Information

**Contents of this file**

**Text S1.**

Consider Rayleigh waves propagating in a horizontally layered model consisting of $N$ layers, as shown in Figure S1. For the $n$th layer, $\alpha_n$, $\beta_n$, $\rho_n$ and $h_n$ represent the P-wave velocity, S-wave velocity, density and thickness, respectively. At the top of the layer, $r_{n,1}^{(t)}$, $r_{n,2}^{(t)}$, $r_{n,3}^{(t)}$ and $r_{n,4}^{(t)}$ represent 4 displacement-stress amplitude coefficients. These 4 coefficients characterize the displacements and stresses at the top of the layer as follows:

$$\begin{cases} u_n^{(t)} = r_{n,1}^{(t)} e^{i(kx-\omega t)} \\ w_n^{(t)} = i r_{n,2}^{(t)} e^{i(kx-\omega t)} \\ t_{zx,n}^{(t)} = r_{n,3}^{(t)} e^{i(kx-\omega t)} \\ \sigma_{zz,n}^{(t)} = i r_{n,4}^{(t)} e^{i(kx-\omega t)} \end{cases} \quad (S1)$$

In Equation S1, $u_n^{(t)}$ and $w_n^{(t)}$ denote the horizontal (x-direction) and vertical (z-direction) displacement components while $\sigma_{zz,n}^{(t)}$ and $t_{zx,n}^{(t)}$ denote the normal and shear stresses acting on the z-plane. $k$ and $\omega$ denote the wave number and angular frequency of the Rayleigh wave. Similarly, $r_{n,1}^{(b)}$, $r_{n,2}^{(b)}$, $r_{n,3}^{(b)}$ and $r_{n,4}^{(b)}$ are the corresponding displacement-stress amplitude coefficients that characterize the displacements and stresses at the bottom of the layer.

A 4×4 propagator matrix $G_n$ maps the motion-stress vector containing the 4 amplitude coefficients at the top of the layer to that at the bottom of the layer (Aki & Richards, 2009):

$$\left[r_{n,1}^{(b)}, r_{n,2}^{(b)}, r_{n,3}^{(b)}, r_{n,4}^{(b)}\right]^T = G_n \left[r_{n,1}^{(t)}, r_{n,2}^{(t)}, r_{n,3}^{(t)}, r_{n,4}^{(t)}\right]^T \quad (S2)$$

$G_n$ can be partitioned into four 2×2 submatrices as follows:

$$G_n = \begin{bmatrix} \left(\frac{\beta_n}{c}\right)^2 A_{11n} & \frac{1}{2\pi f c \rho_n} A_{12n} \\ 2\pi f c \rho_n \left(\frac{\beta_n}{c}\right)^4 A_{21n} & \left(\frac{\beta_n}{c}\right)^2 A_{22n} \end{bmatrix} \quad (S3)$$

Using the propagator matrix method (Gilbert & Backus, 2012), Aki and Richards (Aki & Richards, 2009) derived the elements of all submatrices as follows:

$$A_{11n} = \begin{bmatrix} 2\cosh p_n - (1+v_n^2)\cosh q_n & \frac{(1+v_n^2)}{\gamma_n}\sinh p_n - 2v_n \sinh q_n \\ -2\gamma_n \sinh p_n + \frac{(1+v_n^2)}{v_n}\sinh q_n & -(1+v_n^2)\cosh p_n + 2\cosh q_n \end{bmatrix} \quad (S4)$$

$$A_{12n} = \begin{bmatrix} \dfrac{1}{\gamma_n}\sinh p_n - v_n \sinh q_n & \cosh p_n - \cosh q_n \\ -\cosh p_n + \cosh q_n & -\gamma_n \sinh p_n + \dfrac{1}{v_n}\sinh q_n \end{bmatrix} \quad (S5)$$

$$A_{21n} = \begin{bmatrix} 4\gamma_n \sinh p_n - \dfrac{(1+v_n^2)^2}{v_n}\sinh q_n & 2(1+v_n^2)(\cosh p_n - \cosh q_n) \\ 2(1+v_n^2)(-\cosh p_n + \cosh q_n) & -\dfrac{(1+v_n^2)^2}{\gamma_n}\sinh p_n + 4v_n \sinh q_n \end{bmatrix} \quad (S6)$$

$$A_{22n} = \begin{bmatrix} 2\cosh p_n - (1+v_n^2)\cosh q_n & 2\gamma_n \sinh p_n - \dfrac{(1+v_n^2)}{v_n}\sinh q_n \\ -\dfrac{(1+v_n^2)}{\gamma_n}\sinh p_n + 2v_n \sinh q_n & -(1+v_n^2)\cosh p_n + 2\cosh q_n \end{bmatrix} \quad (S7)$$

In Equations S3 to S7, $f$ and $c$ denote Rayleigh wave frequency and phase velocity. $\gamma_n = \sqrt{1 - \dfrac{c^2}{\alpha_n^2}}$, $v_n = \sqrt{1 - \dfrac{c^2}{\beta_n^2}}$, $p_n = \dfrac{2\pi f h_n \gamma_n}{c}$ and $q_n = \dfrac{2\pi f h_n v_n}{c}$ are 4 dimensionless coefficients. The continuity conditions at layer interfaces are as follows:

$$\left[r_{n+1,1}^{(t)}, r_{n+1,2}^{(t)}, r_{n+1,3}^{(t)}, r_{n+1,4}^{(t)}\right]^T = \left[r_{n,1}^{(b)}, r_{n,2}^{(b)}, r_{n,3}^{(b)}, r_{n,4}^{(b)}\right]^T \quad (S8)$$

By combining Equations S2 and S8, the amplitude coefficients at surface can be mapped to those at the top of the half-space:

$$\left[r_{N,1}^{(t)}, r_{N,2}^{(t)}, r_{N,3}^{(t)}, r_{N,4}^{(t)}\right]^T = G_{N-1} \dots G_2 G_1 \left[r_{1,1}^{(t)}, r_{1,2}^{(t)}, r_{1,3}^{(t)}, r_{1,4}^{(t)}\right]^T \quad (S9)$$

Let $\grave{P}, \acute{P}, \grave{S}$ and $\acute{S}$ denote the displacement amplitudes of downgoing and upgoing P-waves and S-waves in the half-space. Their relationships with the amplitude coefficients at top of the half-space are as follows:

$$\left[\grave{P}, \grave{S}, \acute{P}, \acute{S}\right]^T = \Lambda \bar{T}\left[r_{N,1}^{(t)}, r_{N,2}^{(t)}, r_{N,3}^{(t)}, r_{N,4}^{(t)}\right]^T \quad (S10)$$

In equation S10, $\Lambda = \mathrm{diag}(e^{z\gamma_N}, e^{zv_N}, e^{-z\gamma_N}, e^{-zv_N})$ is a diagonal matrix in which $z$ is the depth at the top of the half-space. The elements of $\bar{T} = T^{-1}$ are as follows:

$$\bar{T} = \begin{bmatrix} \dfrac{\beta_N^2}{\alpha_N c} & -\dfrac{\beta_N^2(1+v_n^2)}{2\alpha_N c \gamma_n} & -\dfrac{1}{4\pi f \rho_N \alpha_N \gamma_n} & \dfrac{1}{4\pi f \rho_N \alpha_N} \\ -\dfrac{\beta_N(1+v_n^2)}{2c v_n} & \dfrac{\beta_N}{c} & \dfrac{1}{4\pi f \rho_N \beta_N} & -\dfrac{1}{4\pi f \rho_N \beta_N v_n} \\ \dfrac{\beta_N^2}{\alpha_N c} & \dfrac{\beta_N^2(1+v_n^2)}{2\alpha_N c \gamma_n} & \dfrac{1}{4\pi f \rho_N \alpha_N \gamma_n} & \dfrac{1}{4\pi f \rho_N \alpha_N} \\ -\dfrac{\beta_N(1+v_n^2)}{2c v_n} & -\dfrac{\beta_N}{c} & -\dfrac{1}{4\pi f \rho_N \beta_N} & -\dfrac{1}{4\pi f \rho_N \beta_N v_n} \end{bmatrix} \quad (S11)$$

The boundary conditions demand zero stresses at the free surface ($r_{1,3}^{(t)} = r_{1,4}^{(t)} = 0$), while the radiation conditions specify that no waves propagate upward in the half-space ($\acute{P} = \acute{S} = 0$).

Substituting Equation S10 into Equation S9 and applying the boundary and radiation conditions yields:

$$[\dot{P}, \dot{S}, 0, 0]^T = B\left[r_{1,1}^{(t)}, r_{1,2}^{(t)}, 0, 0\right]^T \tag{S12}$$

where $B = \Lambda T^{-1} G_{N-1} \ldots G_2 G_1$ is the layer stack matrix. For Equation S12 to hold, the four elements in the lower-left corner of matrix $B$ must satisfy the following condition:

$$b_{31}b_{42} - b_{32}b_{41} = 0 \tag{S13}$$

Equation S13 is the dispersion equation that describes the relationship between Rayleigh wave frequency $f$ and phase velocity $c$. A revision of the forward modeling is necessary to illustrate the scaling properties for Rayleigh wave dispersion. First, by defining $\eta_n = \frac{1}{2\pi f c \rho_n} \left(\frac{c}{\beta_n}\right)^2$, Equation S2 can be reformulated as follows:

$$\left[r_{n,1}^{(b)}, r_{n,2}^{(b)}, \eta_n r_{n,3}^{(b)}, \eta_n r_{n,4}^{(b)}\right]^T = \left(\frac{\beta_n}{c}\right)^2 L_n \left[r_{n,1}^{(t)}, r_{n,2}^{(t)}, \eta_n r_{n,3}^{(t)}, \eta_n r_{n,4}^{(t)}\right]^T \tag{S14}$$

In Equation S15, the elements of $L_n$ are as follows:

$$L_n = \begin{bmatrix} A_{11n} & A_{12n} \\ A_{21n} & A_{22n} \end{bmatrix} \tag{S15}$$

To accommodate the revised motion-stress vector, the continuity conditions at layer interfaces are reformulated as follows:

$$\left[r_{n+1,1}^{(t)}, r_{n+1,2}^{(t)}, \eta_{n+1} r_{n+1,3}^{(t)}, \eta_{n+1} r_{n+1,4}^{(t)}\right]^T = R_n \left[r_{n,1}^{(b)}, r_{n,2}^{(b)}, \eta_n r_{n,3}^{(b)}, \eta_n r_{n,4}^{(b)}\right]^T \tag{S16}$$

where $R_n = \text{diag}(1, 1, \frac{\rho_n \beta_n^2}{\rho_{n+1} \beta_{n+1}^2}, \frac{\rho_n \beta_n^2}{\rho_{n+1} \beta_{n+1}^2})$ is a diagonal matrix. Equation S10 is also reformulated as follows:

$$[\dot{P}, \dot{S}, \dot{P}, \dot{S}]^T = \frac{\beta_N}{2c} \Lambda \bar{T}^* \left[r_{N,1}^{(t)}, r_{N,2}^{(t)}, \eta_N r_{N,3}^{(t)}, \eta_N r_{N,4}^{(t)}\right]^T \tag{S17}$$

In Equation S17, the elements of $\bar{T}^*$ are as follows:

$$\bar{T}^* = \begin{bmatrix} \frac{2\beta_N}{\alpha_N} & -\frac{\beta_N(1+\nu_N^2)}{\alpha_N \gamma_N} & -\frac{\beta_N}{\alpha_N \gamma_N} & \frac{\beta_N}{\alpha_N} \\ -\frac{(1+\nu_N^2)}{\nu_N} & 2 & 1 & -\frac{1}{\nu_N} \\ \frac{2\beta_N}{\alpha_N} & \frac{\beta_N(1+\nu_N^2)}{\alpha_N \gamma_N} & \frac{\beta_N}{\alpha_N \gamma_N} & \frac{\beta_N}{\alpha_N} \\ -\frac{(1+\nu_N^2)}{\nu_N} & -2 & -1 & -\frac{1}{\nu_N} \end{bmatrix} \tag{S18}$$

By combining Equations S14, S16 and S17, a revised layer stack matrix $B^*$ can be derived as follows:

$$B^* = \frac{\beta_N}{2c} \prod_{n=1}^{N-1} \left(\frac{\beta_n}{c}\right)^2 \Lambda \bar{T}^* R_{N-1} L_{N-1} \ldots R_2 L_2 R_1 L_1 \tag{S19}$$

The dispersion equation remains the same for $B^*$, which is $b_{31}^* b_{42}^* - b_{32}^* b_{41}^* = 0$. By defining $B' = \bar{T}^* R_{N-1} L_{N-1} \ldots R_2 L_2 R_1 L_1$, the dispersion equation can be rewritten as:

$$b_{31}^*b_{42}^* - b_{32}^*b_{41}^* = \frac{\beta_N}{2c}\prod_{n=1}^{N-1}\left(\frac{\beta_n}{c}\right)^2 e^{-z\gamma_N}e^{-zv_N}(b'_{31}b'_{42} - b'_{32}b'_{41}) = 0 \qquad (S20)$$

Equation S20 shows that the coefficient $\frac{\beta_N}{2c}\prod_{n=1}^{N-1}\left(\frac{\beta_n}{c}\right)^2$ and matrix $\Lambda$ have no influence on the dispersion equation and therefore can be removed from the computation. The last step of reformulation involves removing the first 2 rows of matrix $\bar{T}^*$ as they are not involved in the computation of elements $b'_{31}$, $b'_{32}$, $b'_{41}$ and $b'_{42}$. The final layer stack matrix $D$ is computed as follows:

$$D = SR_{N-1}L_{N-1}\ldots R_2L_2R_1L_1 \qquad (S21)$$

where $S$ is a $2\times 4$ matrix comprising the last 2 rows of $\bar{T}^*$. The corresponding dispersion equation is:

$$F(f,c) = d_{11}d_{22} - d_{12}d_{21} = 0 \qquad (S22)$$

For the horizontally layered model shown in Figure S1, the dispersion equation $F(f,c)$ is governed by three categories of layer parameters: mass density $\rho_n$, layer thickness $h_n$, and elastic wave velocities $\alpha_n$ and $\beta_n$. For each parameter category, there exists a corresponding scaling property that characterizes the transformation of the dispersion equation $F(f,c)$ when the parameters within that category are scaled by the same factor across all layers.

The density scaling property states that the dispersion equation $F(f,c) = 0$ remains invariant when mass density $\rho_n$ are scaled by a common factor across all layers. This invariance arises because density parameters appear only as ratios $\frac{\rho_n}{\rho_{n+1}}$ in matrix $R_n$ during the computation of the layer stack matrix $D$. Since the density ratios $\frac{\rho_n}{\rho_{n+1}}$ are preserved under uniform scaling, the matrix $R_n$ and consequently the layer stack matrix $D$ remain unchanged, indicating that the original and scaled models yield identical dispersion curves.

The thickness scaling property demonstrates that when all layer thicknesses $h_n$ are scaled by a common factor $\lambda$, the dispersion equation transforms from $F(f,c) = 0$ to $F(\lambda f,c) = 0$. This transformation implies that if $(f,c)$ is a point on the original dispersion curve, then $\left(\frac{f}{\lambda},c\right)$ must correspondingly be a point on the dispersion curve of the scaled model. This scaling relationship can be explained by examining two parameters $p_n = \frac{2\pi f h_n \gamma_n}{c}$ and $q_n = \frac{2\pi f h_n v_n}{c}$, which are the only parameters affected by $h_n$ and $f$ during the computation of $D$. For the scaled model, $h_n$ is scaled by $\lambda$. If $f$ is scaled by $\frac{1}{\lambda}$ simultaneously, $p_n$, $q_n$ and consequently $D$ remain invariant, confirming that $F(\lambda f,c) = 0$ must be the dispersion equation of the scaled model.

Similarly, the velocity scaling property reveals that when all elastic wave velocities $\alpha_n$ and $\beta_n$ are scaled by a common factor $\lambda$, the dispersion equation transforms from $F(f,c) = 0$ to $F\left(\frac{f}{\lambda},\frac{c}{\lambda}\right) = 0$. It suggests that if $(f,c)$ is a point on the original dispersion curve, then $(\lambda f,\lambda c)$ must be a point on the dispersion curve of the scaled model. In the scaled model, $c$ and $f$ are scaled by $\lambda$ to ensure that the 4 dimensionless coefficients $\gamma_n$, $v_n$, $p_n$ and $q_n$ maintain original values when $\alpha_n$ and $\beta_n$ are scaled by the same factor. As a result, elements in matrices $S$, $R_n$ and $L_n$ remain unchanged, validating that $(\lambda f, \lambda c)$ satisfies the dispersion equation of the scaled model.

**Text S2.**

Generating S-wave velocity profiles for the 100 thin layers while maintaining the no-LVZ constraint requires careful consideration. A critical consideration in generating training data for deep learning is avoiding bias. For instance, if the S-wave velocity of the top 5 meters is consistently around 500 m/s across all training samples, the model would invariably learn the pattern and output similar values during inversion, which does not reflect real-world variability. Therefore, the random generation algorithm must be designed to ensure that each layer has a uniform probability of attaining any velocity between 50 and 1000 m/s, where the upper bound corresponds to the half-space S-wave velocity. The lower bound is set as 1/20 (50 m/s) of the half-space S-wave velocity, as values below this threshold are very uncommon in measured dispersion curves due to the trade-off between resolution and penetration depth.

A straightforward approach to generate S-wave velocity for each layer while maintaining the no-LVZ constraint is the Top-down Method, in which the S-wave velocity for each layer is generated sequentially from top to bottom layers. To enforce the no-LVZ constraint, the S-wave velocity for each layer is uniformly sampled between the S-wave velocity of the layer immediately above and the S-wave velocity for the half-space.

Ten thousand S-wave velocity profiles were generated following the Top-down Method, with the statistical distribution of S-wave velocities for each layer visualized in Figure S2(a). This figure presents a contour plot depicting the distribution of velocity values as a function of depth for S-wave velocity profiles generated following the Top-down Method. The horizontal axis represents S-wave velocity, ranging from 50 to 1000 m/s, segmented into intervals of 50 m/s each, while the vertical axis denotes depth. The frequency with which S-wave velocities fall within each velocity interval at various depths is represented through a color gradient scheme, where different colors indicate varying frequencies. As shown in Figure S2(a), the S-wave velocities for most layers are concentrated near 1000 m/s, except in the top few layers. It indicates that the Top-down Method introduces a strong bias and is therefore unsuitable for generating the S-wave velocity profiles for machine learning.

Wathelet proposed a method for generating S-wave velocity profiles that complies with the no-LVZ constraint, termed the Diagonal Method (Wathelet, 2005). This method begins by uniformly sampling the S-wave velocities of the first and last layers between 50 and 1000 m/s. Then, considering a conventional plot of S-wave velocity profiles where the vertical axis represents depth and the horizontal axis represents velocities, the method draws a diagonal line from the bottom left to the top right of the plot. A point is randomly selected along this diagonal line, which divides the plot into two distinct sections. This process is then performed recursively on both resulting sections. Ultimately, all the selected points collectively constitute the S-wave velocity profile. Figure S2(b) illustrates the statistical distribution of S-wave velocities for each layer across ten thousand S-wave velocity profiles generated using the Diagonal Method. The distribution is generally uniform throughout the plot, with the exception of a diagonal band extending from the top left to bottom right. This diagonal band indicates that S-wave velocities are more likely to fall within specific velocity ranges at corresponding depths, and these velocity ranges increase linearly with depth. Figure S2(c) shows the statistical distribution of S-wave velocities for each layer across ten thousand S-wave velocity profiles generated using the proposed Randomized Layer Assignment Method. The frequency distribution is very uniform, with only a slight concentration in the top left and bottom right regions. Overall, the Randomized Layer Assignment Method shows the least bias among all three methods. Therefore, training and testing data were generated using this method.

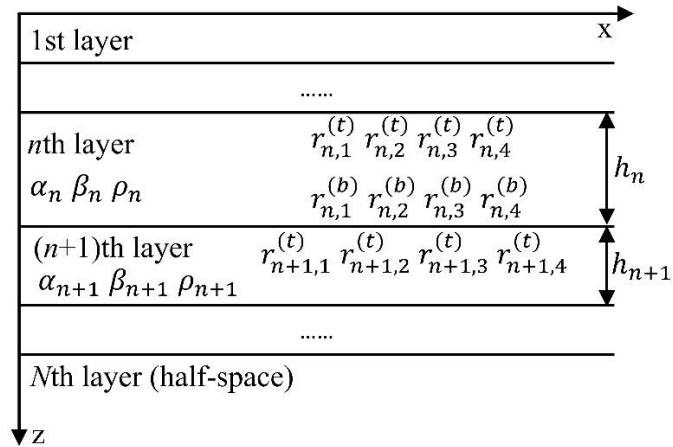

**Figure S1.** Schematic of a horizontally layered model consisting of N layers.

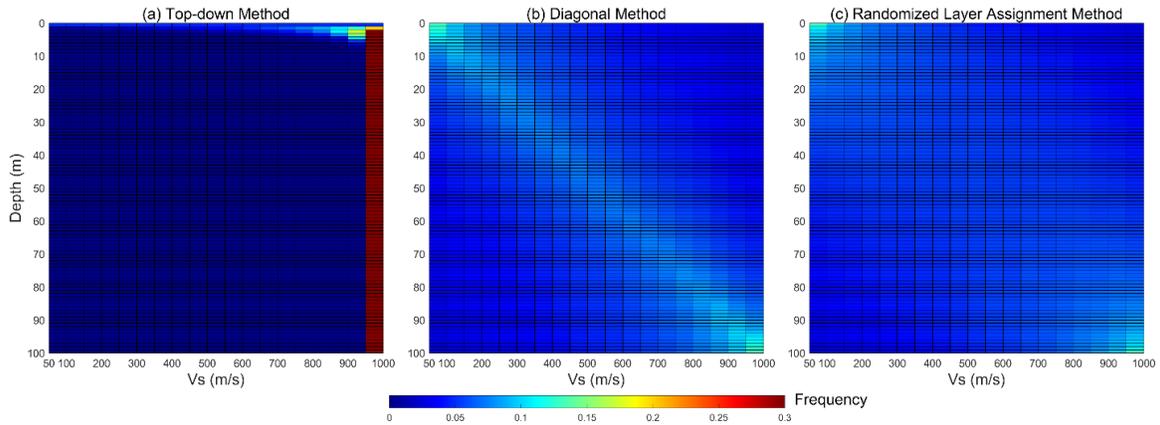

**Figure S2.** Histograms of S-wave velocity distribution at different depths for ten thousand randomly generated profiles using: (a) Top-down Method, (b) Diagonal Method and (c) Randomized Layer Assignment Method. At each depth, the color of the grids indicates the probability that the S-wave velocity falls within the corresponding velocity range.

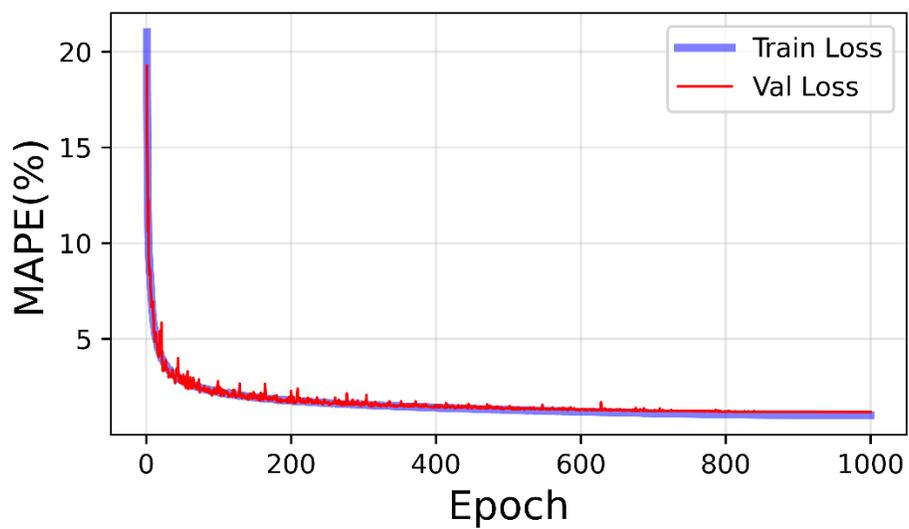

**Figure S3.** The training loss and validation loss curves of the model over 1000 epochs, measured in Mean Absolute Percentage Error (MAPE).

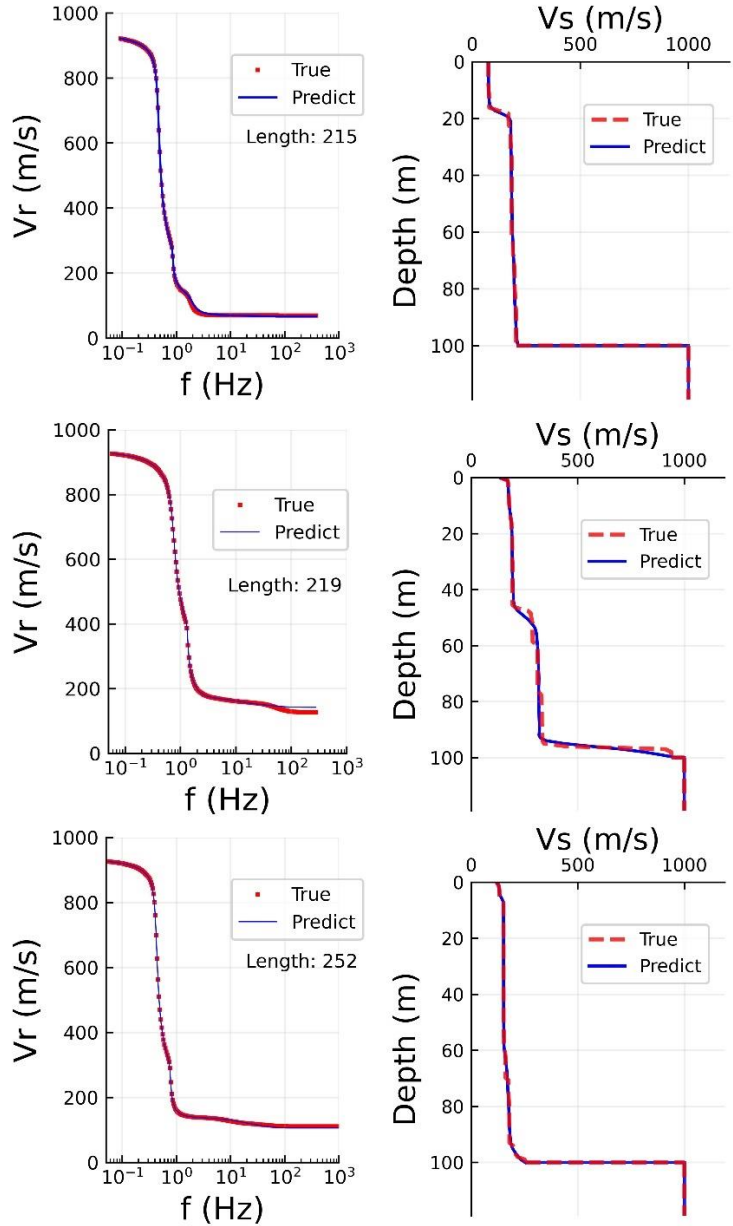

**Figure S4.** Comparison between true data and predicted results from the trained model for three examples of long dispersion curves. The left panels display the input dispersion curves, and the right panels display the corresponding inverted $V_S$ profiles. The 'Length' specifies the number of input points from each curve fed into the model.

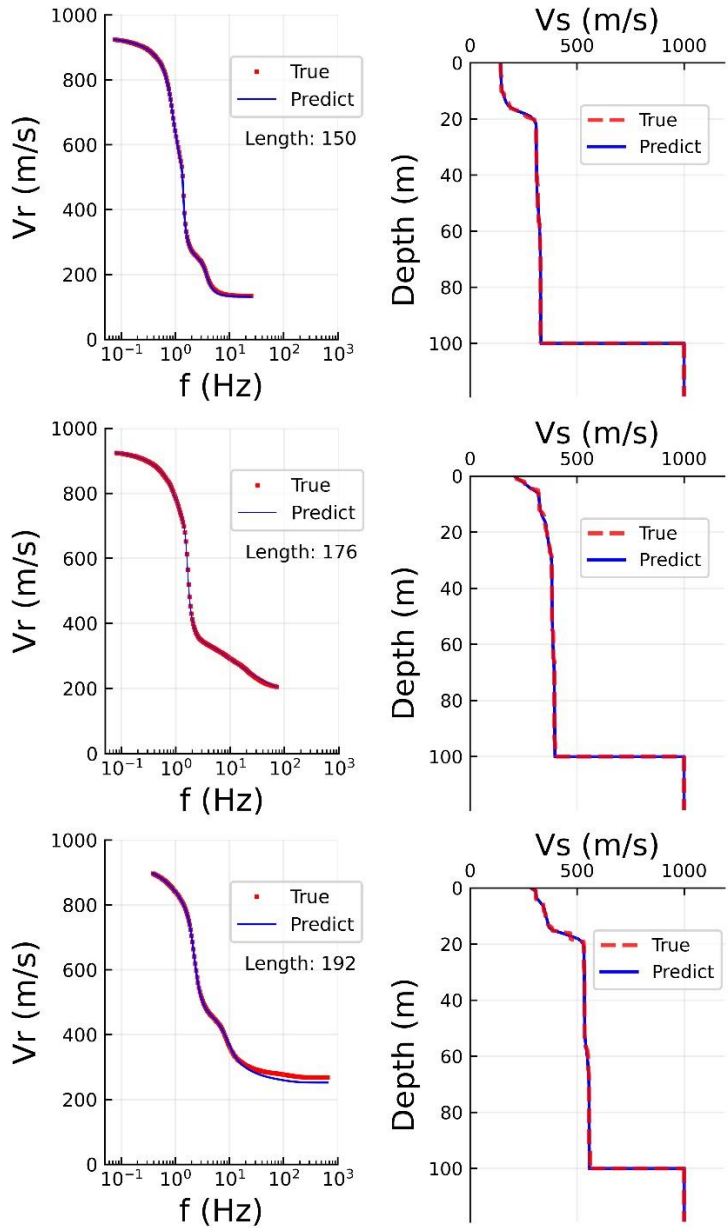

**Figure S5.** Comparison between true data and predicted results from the trained model for three examples of medium-length dispersion curves. The left panels display the input dispersion curves, and the right panels display the corresponding inverted $V_S$ profiles. The 'Length' specifies the number of input points from each curve fed into the model.

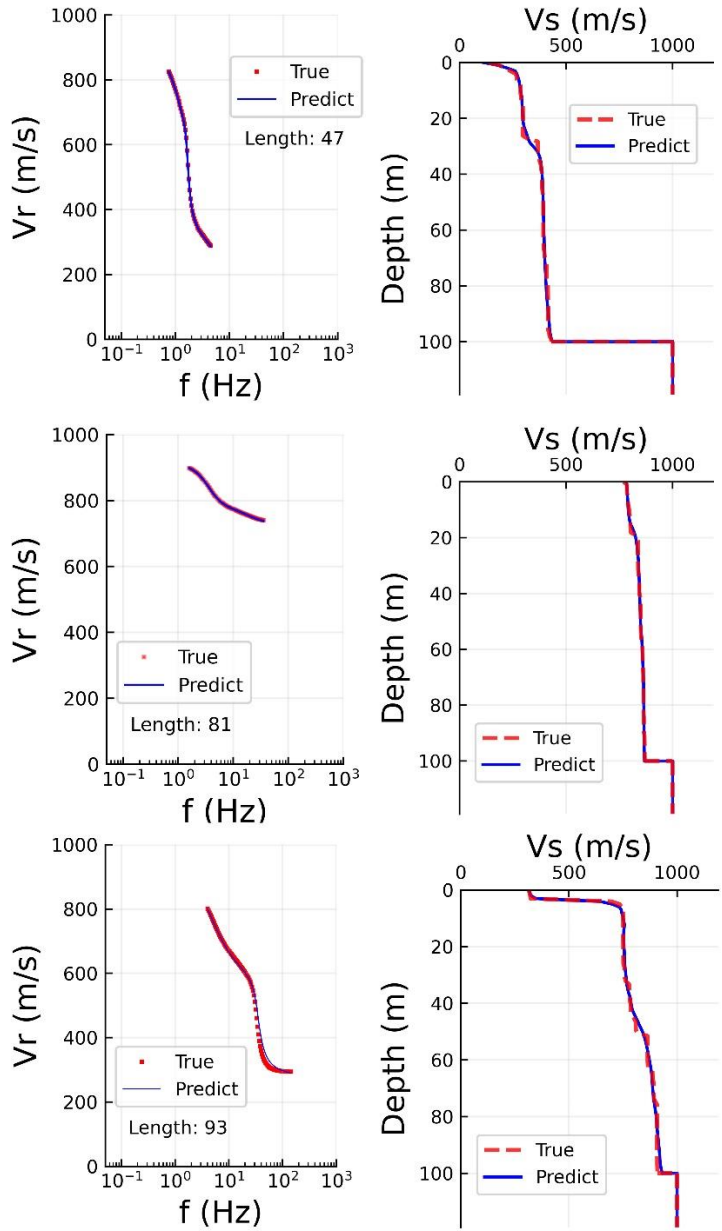

**Figure S6.** Comparison between true data and predicted results from the trained model for three examples of short dispersion curves. The left panels display the input dispersion curves, and the right panels display the corresponding inverted $V_S$ profiles. The 'Length' specifies the number of input points from each curve fed into the model.

**Table S1.** Layer parameters for the base model.

| Layer number | β (m/s) | α (m/s) | ρ (g/cm³) | h (m) |
|---|---|---|---|---|
| 1 | 194 | 650 | 1.82 | 2 |
| 2 | 270 | 750 | 1.86 | 2.3 |
| 3 | 367 | 1400 | 1.91 | 2.5 |
| 4 | 485 | 1800 | 1.96 | 2.8 |
| 5 | 603 | 2150 | 2.02 | 3.2 |
| Half-space | 740 | 2800 | 2.09 | Infinite |

**Table S2.** Layer parameters for the velocity-scaled model.

| Layer number | β (m/s) | α (m/s) | ρ (g/cm³) | h (m) |
|---|---|---|---|---|
| 1 | 291 | 975 | 1.82 | 2 |
| 2 | 405 | 1125 | 1.86 | 2.3 |
| 3 | 550.5 | 2100 | 1.91 | 2.5 |
| 4 | 727.5 | 2700 | 1.96 | 2.8 |
| 5 | 904.5 | 3225 | 2.02 | 3.2 |
| Half-space | 1100 | 4200 | 2.09 | Infinite |

**Table S3.** Layer parameters for the thickness-scaled model.

| Layer number | β (m/s) | α (m/s) | ρ (g/cm³) | h (m) |
|---|---|---|---|---|
| 1 | 194 | 650 | 1.82 | 3 |
| 2 | 270 | 750 | 1.86 | 3.45 |
| 3 | 367 | 1400 | 1.91 | 3.75 |
| 4 | 485 | 1800 | 1.96 | 4.2 |
| 5 | 603 | 2150 | 2.02 | 4.8 |
| Half-space | 740 | 2800 | 2.09 | Infinite |

**Table S4.** Layer parameters for the density-scaled model.

| Layer number | β (m/s) | α (m/s) | ρ (g/cm³) | h (m) |
|---|---|---|---|---|
| 1 | 194 | 650 | 2.184 | 2 |
| 2 | 270 | 750 | 2.232 | 2.3 |
| 3 | 367 | 1400 | 2.292 | 2.5 |
| 4 | 485 | 1800 | 2.352 | 2.8 |
| 5 | 603 | 2150 | 2.424 | 3.2 |
| Half-space | 740 | 2800 | 2.508 | Infinite |